\documentclass[twocolumn,amsmath,amssymb,superscriptaddress,aps,pra]{revtex4-1}

\usepackage{color}% color text
\usepackage{graphicx}
\usepackage{amssymb}
\usepackage{amsmath}
\usepackage{mathrsfs}
\usepackage{supertabular}
\usepackage{makecell}
\newcommand{\mynomencl}[3][subsection]{%
  \begingroup\edef\x{\endgroup
  \unexpanded{\nomenclature{#2}}%
    {\unexpanded{#3} (\csname the#1\endcsname)}}\x}

\newcommand{\ud}{\mathrm{d}}

\newcommand{\pdern}[3]{\frac{\partial^#3#1}{\partial#2^#3}}
\newcommand{\pder}[2]{\frac{\partial#1}{\partial#2}}
\let\Eps\varepsilon
\newcommand{\Deltalpha} {\Delta\!\alpha}
\usepackage[breaklinks=true]{hyperref}
\graphicspath{ {./figures/} {./}}

\begin{document}

\title{Stabilization of uni-directional water wave trains over an uneven bottom}

\author{Andrea Armaroli} 

\affiliation{Institute for Environmental Sciences, Universit{\'e} de Gen{\`e}ve, Boulevard Carl-Vogt 66, 1205 Gen{\`e}ve, Switzerland }  \affiliation{GAP, Universit{\'e} de Gen{\`e}ve, Chemin de Pinchat 22, 1227 Carouge, Switzerland}

\author{Alexis Gomel} 
\affiliation{Institute for Environmental Sciences, Universit{\'e} de Gen{\`e}ve, Boulevard Carl-Vogt 66, 1205 Gen{\`e}ve, Switzerland }  \affiliation{GAP, Universit{\'e} de Gen{\`e}ve, Chemin de Pinchat 22, 1227 Carouge, Switzerland}

\author{
Amin Chabchoub } 
\affiliation{ Centre for Wind, Waves and Water, School of Civil Engineering, The University of Sydney, Sydney, NSW 2006, Australia}
\affiliation{Marine Studies Institute, The University of Sydney, Sydney, NSW 2006, Australia
}

\author{Maura Brunetti} 
\affiliation{Institute for Environmental Sciences, Universit{\'e} de Gen{\`e}ve, Boulevard Carl-Vogt 66, 1205 Gen{\`e}ve, Switzerland }  \affiliation{GAP, Universit{\'e} de Gen{\`e}ve, Chemin de Pinchat 22, 1227 Carouge, Switzerland}
\author{J{\'e}r{\^o}me Kasparian} 
\affiliation{Institute for Environmental Sciences, Universit{\'e} de Gen{\`e}ve, Boulevard Carl-Vogt 66, 1205 Gen{\`e}ve, Switzerland }  \affiliation{GAP, Universit{\'e} de Gen{\`e}ve, Chemin de Pinchat 22, 1227 Carouge, Switzerland}

\begin{abstract}
We study the evolution of nonlinear surface gravity water wave packets developing from modulational instability over an uneven bottom. A nonlinear Schrödinger equation (NLSE) with coefficients varying in space along propagation is used as a reference model. Based on a low-dimensional approximation obtained by considering only three complex harmonic modes, we discuss how to stabilize a one-dimensional pattern in the form of train of large peaks sitting on a background and propagating over a significant distance. Our approach is based on a gradual depth variation, while its conceptual framework is the theory of autoresonance in nonlinear systems and leads to a quasi-frozen state. Three main stages are identified: amplification from small sideband amplitudes, separatrix crossing, and adiabatic conversion to orbits oscillating around an elliptic fixed point. Analytical estimates on the three stages are obtained from the low-dimensional approximation and validated by NLSE simulations. Our result will contribute to understand the dynamical stabilization of nonlinear wave packets and the persistence of large undulatory events in hydrodynamics and other nonlinear dispersive media.\end{abstract}

\maketitle

\section*{List of symbols}

\begin{supertabular}{c l l} 
 {$\omega$} &  Angular frequency in m\textsuperscript {$-\frac {1}{2}$} & Sec.~\ref{ssec:VDNLS}\\
 {$\kappa $} & Wavenumber-depth product (ad.) & Sec.~\ref{ssec:VDNLS}\\
  $\Eps $ & Wave steepness & Sec.~\ref{ssec:VDNLS}  \\
  {$\sigma $} & Depth correction factor & Sec.~\ref{ssec:VDNLS}\\
 $c_\mathrm {g}$ & Group velocity& Sec.~\ref{ssec:VDNLS}\\

{$\beta $} & Dispersion coefficient & App.~\ref{app:VDcoeffs}\\
$\gamma $ &  Nonlinear coefficient & App.~\ref{app:VDcoeffs}\\ 

$\tilde \gamma $ & Effective nonlinear coefficient & Sec.~\ref{ssec:VDNLS}\\
{$\mu _0$} &  Shoaling coefficient & App.~\ref{app:VDcoeffs}\\
   
$U(\xi ,\tau )$ & \makecell[l]{ Complex envelope of \\surface elevation} & Sec.~\ref{ssec:VDNLS} \\
$V(\xi ,\tau )$ & Shoaling-corrected $U$ & Sec.~\ref{ssec:VDNLS}\\

$V_0$ & Carrier amplitude & Sec.~\ref{ssec:MI}\\
    
$\alpha $ &  Three-wave parameter & Sec.~\ref{ssec:NLMI}\\

$\Omega $ &  Modulation detuning &Sec.~\ref{ssec:NLMI}\\

$\Omega _\mathrm {C}$ &  Cut-off MI detuning & Sec.~\ref{ssec:MI}  \\

$\Omega _\mathrm {M}$ & Peak MI detuning & Sec.~\ref{ssec:MI} \\

$\psi $ & Relative sideband phase & Sec.~\ref{ssec:NLMI}\\

$\eta $ & Conversion rate to sidebands & Sec.~\ref{ssec:NLMI} \\

$(\tilde \psi _i,\tilde \eta _i)$ & \makecell[l]{Fixed points of \\ the three-wave system} & Sec.~\ref{ssec:NLMI}\\

$H^{(\xi )}$ & \makecell[l]{Hamiltonian function of \\the three-wave system} & Sec.~\ref{ssec:NLMI}\\

$H^{(X)}$ & \makecell[l] {Hamiltonian function of\\ the three-wave system\\ (w.r.t. the reduced variable $X$)} & Sec.~\ref{ssec:NLMI}\\

$H_\mathrm{min}$ & \makecell[l]{Value of the Hamiltonian \\ function at centers} & Sec.~\ref{ssec:NLMI} \\

$\Deltalpha ^\mathrm {i}$ & \makecell[l]{Slope of $\alpha $\\ in the linear stage} & Sec.~\ref{ssec:linear}\\

$\Deltalpha ^\mathrm {t}$ &\makecell[l]{Slope of $\alpha $\\ in the intermediate stage} & Sec.~\ref{ssec:interm}\\

$\Deltalpha ^\mathrm {f}$ &\makecell[l]{Slope of $\alpha $\\ in the adiabatic stage} & Sec.~\ref{ssec:adiab}\\
$X^*$ & Crossing into MI sideband  & Sec.~\ref{ssec:linear}\\

$X^{**}$ &  Start of adiabatic regime  & Sec.~\ref{ssec:interm}\\

\end{supertabular}

% \printnomenclature

\section{Introduction}

Modulational instability (MI) is an ubiquitous phenome-non for wave packets propagating in a weakly nonlinear medium \cite{Benjamin1967}. It consists in the appearance of sidebands growing around a uniformly-modulated carrier and was observed  in deep water waves, nonlinear optics, Bose-Einstein condensates, and plasma physics \cite{Zakharov2009,Dudley2019}. 

If the envelope of the wave-packet is  narrowbanded, the nonlinear stage of the evolution (\textit{i.e.}, when the sidebands start to grow at amplitudes comparable to the unstable stationary background) can be modeled by means of the universal nonlinear Schr{\"o}dinger equation (NLSE). This integrable equation exhibits  exact solutions, e.g., stationary envelope solitons and pulsating breathers of Kuznetsov-Ma-, Peregrine- and Akhmediev-type \cite{Kuznetsov1977,Ma1979,Peregrine1983,Akhmediev1986}

The family of Akhmediev breather (AB) is the prototype of the nonlinear evolution of MI: in the time-like NLSE, an initially slightly modulated time-periodic train of pulses reaches its peak value at a given point in space, as a result of the exponential sideband growth, as is followed by the recovery of the initial state known as Fermi-Pasta-Ulam recurrence \cite{Akhmediev2001e}. Because of this characteristic feature, \textit{i.e.},
extreme waves appearing from nowhere and suddenly disappearing
\cite{Akhmediev2009a}, it is also a candidate solution for the explanation of rogue waves and other nonlinear systems. 

The NLSE is a framework not only valid for deep water, but also for intermediate depth cases, as is well known from the literature \cite{Hasimoto1972,Davey1974,Chiang2005}.

The depth is thus an important degree of freedom that tunes the dispersion  and nonlinear coefficients during wave propagation, thus  allowing the possibility to dynamically control the MI gain. In optics, an adiabatic variation of fiber dispersion is well known to provide an effective path to soliton compression \cite{Smith1989,Chernikov1991}. Moreover, the transition from two  fibers of different dispersion was recently proposed to control an AB at its peak focusing point \cite{Bendahmane2014a}. A standard fiber has a large cross-section, thus the nonlinear coefficient does not change much (as it depends mostly on the core area and Kerr nonlinear refractive index); the dispersion is instead much more sensitive to geometry \cite{Agrawal2012}. The opposite is true for surface gravity waves in water: the group velocity always decreases with frequency increase, while the nonlinearity can be tuned to positive or negative values \cite{Chabchoub2015b}. 

Here, we propose a theoretical framework for the control of breathing water wave-packets over a smoothly varying uneven bottom. 
A three-wave truncation \cite{Trillo1991c,Armaroli2017} allows us to formulate the conditions required for stabilization, as well as the limits of our approach. 

We rely on a mechanism similar to autoresonance, in which the change of an external parameter in the system  allows one to lock it in a stable and stationary oscillating state of large amplitude, starting from an initial condition close but not exactly matching the resonant condition. This theory finds its origin in accelerator and plasma physics \cite{Golovanivsky1980,Friedland1992a,Friedland1992} and was recently applied also to optical frequency conversion \cite{Aranson1992,Yaakobi2010c,Yaakobi2013c,Suchowski2014}.

Sec.~\ref{sec:model} recalls the generalized NLSE model in finite water depth of Ref.~\cite{Djordjevic1978} and the description of the nonlinear stage of MI by means of a three-wave truncation approach. In Sec.~\ref{sec:adiab} we discuss the conditions for stabilization and report improvements on the implementation of the abrupt transition as proposed in \cite{Bendahmane2014a}. Numerical results are presented in Sec.~\ref{sec:results}. Sec.~\ref{sec:concl} is devoted to result summary and outlook.

\section{Model equation}
\label{sec:model}
\subsection{Generalized finite water depth NLSE}
\label{ssec:VDNLS}

%%%
% Scheme of the wave system
\begin{figure}[tb]
\centering
\includegraphics[width=0.45\textwidth]{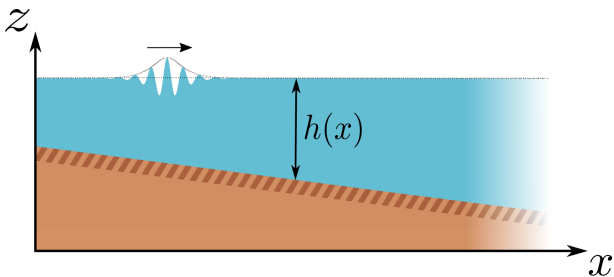}
\caption{ \color{black} Schematic representation of a surface wave packet propagating over an uneven bottom. The propagation is uni-directional from left to right. Wavelengths, depth and amplitude are not in real scales.}
\label{fig:scheme}
\end{figure}

In \cite{Djordjevic1978}, a NLSE-like equation is derived for the one-dimensional evolution of the envelope of surface water waves on an uneven bottom of depth $h$ at frequency $\omega=\sqrt{gk\sigma}$, with $\sigma\equiv\tanh{\kappa}$ and $\kappa\equiv k h$, $k$ being the local wavenumber, which varies with $h$, while $\omega$ is fixed.

The 2D Laplace equation $\left[\pdern{}{x}{2}+\pdern{}{z}{2}\right]\Phi=0$  for the velocity potential $\Phi$ in the longitudinal and depth coordinates $(x,z)$ is solved with the usual kinematic and dynamic boundary conditions at the free surface \cite{Chiang2005}, whereas the bottom boundary condition reads as
\begin{equation}
	\pder{\Phi}{z}=-h'(x)\pder{\Phi}{x}, \,\, z=-h(x).
	\label{eq:bottomBC}
\end{equation}			
	It is required that the bottom slope is small enough to prevent wave-reflections due to wavenumber mismatches, \textit{i.e.}, $h'(x)=\mathcal{O}( \Eps^2)$, with $\Eps\equiv k a\ll 1$ the wave steepness ($a$ is the local carrier wave amplitude). 	
	
By employing the standard method of multiple scales up to third-order in $\Eps$ \cite{Chiang2005}, the following evolution equation was derived \cite{Djordjevic1978}
\begin{equation}
i\pder{U}{\xi} + \beta\pdern{U}{\tau}{2} - \gamma|U|^2U = -i \mu U -i \nu U,
\label{eq:DReq1}
\end{equation}
where  \textcolor{black}{$U(\xi,\tau)$ is the envelope of the free-surface water elevation}, with $\xi\equiv\Eps^2 x$ and $\tau\equiv \Eps\left[\int_0^x \frac{\ud \zeta}{c_\mathrm{g}(\zeta)}-t\right]$ are the coordinates in a frame moving at the group velocity of the envelope, $c_\mathrm{g} \equiv \pder{\omega}{k}= \frac{g}{2\omega}\left[\sigma + \kappa(1-\sigma^2)\right]$; moreover $\beta$, $\gamma$, and $\mu\equiv\mu_0 \frac{\ud \kappa}{\ud \xi}$ represent the dispersion, cubic nonlinearity and shoaling coefficient, respectively.  The first two are simply the coefficients of the NLSE on arbitrary depth, see \cite{Hasimoto1972}, and are functions of $\kappa$ only; detailed expressions can be found in Appendix \ref{app:VDcoeffs}; $\mu$ results from wave-energy conservation arguments as  $\mu_0\equiv\frac{1}{2{\omega c_g}}\frac{\ud \left[\omega c_g\right]}{\ud \kappa}$, \textit{i.e.},  $\mu$ is the logarithmic derivative of $c_\mathrm{g}$. At variance with \cite{Djordjevic1978}, we include also a homogeneous loss term, $\nu$ due to, \textit{e.g.}, viscosity or friction with bottom and sidewalls, which is appropriate at the NLSE order \cite{Segur2005,Dias2008}.

\begin{figure}[tb]
\centering
\includegraphics[width=0.45\textwidth]{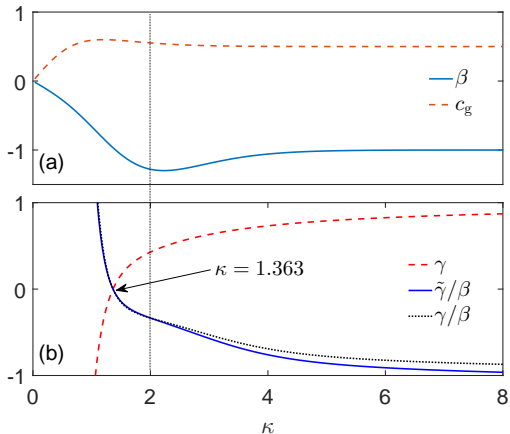}
\caption{Dependence on the depth parameter $\kappa$ of the coefficients of Eq.~\eqref{eq:DReq1}, with $g=\omega=1$ (a) Dispersion parameters: $c_\mathrm{g}$ (dashed red line) and $\beta$ (solid blue line); (b) Nonlinear parameters: $\gamma$ (dashed red line), $\tilde\gamma/\beta$ (solid blue line), and $\gamma/\beta$ (black dotted line),  \textcolor{black}{see Sec.~\ref{ssec:VDNLS} and App.~\ref{app:VDcoeffs}}.  \textcolor{black}{ $\tilde \gamma$ is defined in Eq.~\eqref{eq:effectiveNL}: for definiteness} we take  $\nu=0$ and $\kappa(\xi=0)=2$, marked by a black dotted vertical line. Notice that the impact of shoaling is minor.}
\label{fig:DRparams}
\end{figure}

Let $g=\omega=1$ for definiteness.
It is  well-known that $\beta<0$ for all values of $\kappa$ (provided that only surface gravity waves are considered) [blue solid line in Fig.~\ref{fig:DRparams}(a)], while $\gamma\ge0$ for $\kappa\ge1.363$ [red dashed line in Fig.~\ref{fig:DRparams}(b)].  Recall also that $c_\mathrm{g}$ [red dashed line in Fig.~\ref{fig:DRparams}(a)] is maximum for $\kappa \approx 1.20$.

The form of $\mu$ allows us to simplify Eq.~\eqref{eq:DReq1}.  {Following \cite{Onorato2011}, we let  $U=V\exp\left[-\int_0^\xi{\mu (y) \,\ud y} - \nu\xi\right]$;} Eq.~\eqref{eq:DReq1} can be rewritten as
\begin{equation}
i\pder{V}{\xi} + \beta\pdern{V}{\tau}{2} - \tilde\gamma|V|^2V = 0,
\label{eq:DReq2}
\end{equation}
 \textit{i.e.}, a NLSE with varying parameters, with
 
\begin{equation}
 \tilde\gamma(\xi) \equiv \gamma(\xi) \frac{c_\mathrm{g}(\xi=0)}{c_\mathrm{g}(\xi)}\exp(-2\nu \xi).
 \label{eq:effectiveNL}
\end{equation}  
The effect of shoaling is clear from Eq.~\eqref{eq:effectiveNL}: in the focusing regime, $\beta\tilde\gamma<0$, it slightly increases the effective nonlinearity, because $c_\mathrm{g}$ monotonically decreases, see the red dashed line in Fig.~\ref{fig:DRparams}(a). The effect of $\nu$ is to decrease the impact of nonlinearity as the wave propagates. It is easy to verify that the perfect compensation of $\nu$ by shoaling is impossible for increasing depth. For the sake of simplicity, we will take $\nu=0$ in what follows, except for Sec.~\ref{ssec:numbers}. 
 
In the framework of field theory, Eq.~\eqref{eq:DReq2} conserves the total mass $N\equiv\int_{-\infty}^{\infty}{ |V|^2 \ud \tau}$ and the momentum $P \equiv \mathrm{Im}\left\{\int_{-\infty}^{\infty}{ V^*\pder{V}{\tau}}\ud \tau\right\}$. We use them in our numerical calculations to assure the precision of solutions. No other conserved quantity is present, if coefficients have no specific functional dependence. 	

\subsection{Modulation instability}
\label{ssec:MI}
Eq.~\eqref{eq:DReq2} possesses a steady-state solution $V_s(\xi) = V_0 \allowbreak\exp\left(-iV_0^2\int_0^\xi \tilde\gamma(y)\ud y\right)$. 
For $\beta\tilde\gamma<0$, this solution is modulationally unstable for a detuning $\Omega\in\left[0,\Omega_\mathrm{C}\right]$ from the central frequency $\omega$, with $\Omega_\mathrm{C}\equiv \sqrt{2\left|\frac{\tilde\gamma}{\beta}\right|}V_0$. The linear MI gain is $G = |\beta\Omega|\sqrt{\Omega_\mathrm{C}^2-\Omega^2}$, with peak at $\Omega_\mathrm{M}\equiv\frac{\Omega_\mathrm{C}}{\sqrt{2}}$. This is the result of the conventional linear stability analysis, but it can also be thought of as the nonlinear phase-matching condition between the steady-state solution and the two sidebands, a sort of nonlinear resonance condition. 

% Thus, the ratio $\tilde\gamma/\beta$ is the main parameter of our problem. 
To compute $\tilde\gamma$, the initial value of $\kappa$ must be fixed. As an example, we let $\kappa(0)=2$ and thus $c_g(\xi=0)=0.55$. 
The main parameter of our problem, $\tilde\gamma/\beta$, is shown as blue solid line in Fig.~\ref{fig:DRparams}(b). For comparison, we also include the ratio $\gamma/\beta$, as a dotted black line, to show that the effect of shoaling on MI is quite small (less than 5\%) in the focusing regime. 
As this parameter is changed the same sideband frequency can turn from modulationally stable to unstable or experience a different instability gain along the MI curve. In Fig.~\ref{fig:DRparams}(b), it is clear that the range of variation is quite limited, compared to optical fibers, because both $\beta$ and $\gamma$ tend to their deep-water limits as $\kappa\to\infty$. The choice of the reference value $\kappa=2$ (marked in Fig.~\ref{fig:DRparams}) is a good trade-off for having strong enough nonlinear effects, while avoiding high-order corrections appearing when $\tilde\gamma\approx 0$, see for instance \cite{Sedletsky2003,Slunyaev2005}. 

The MI gain is a linear approximation, beyond which the nonlinear behavior demands a more detailed analysis. 

\subsection{Nonlinear regime}
\label{ssec:NLMI}

A thorough understanding of the problem can come from a low-dimensional analysis. We follow the three-wave truncation proposed in Ref.~\cite{Trillo1991c}, that was proven effective also in higher-order generalizations of the NLSE \cite{Armaroli2017,Armaroli2018}.

Let $V(\xi,\tau)= A_0(\xi) + A_1(\xi) e^{i \Omega \tau} + A_{-1}(\xi)e^{-i\Omega \tau}$, where $\Omega$ is the angular detuning in normalized units, and $A_{n}$, with  {$n\in\{-1,0,1\}$} are complex variables, the phases of which are denoted by $\phi_n$. It is easy to reduce Eq.~\eqref{eq:DReq2} to a one degree-of-freedom (d.o.f.) Hamiltonian system \cite{Trillo1991c}. The canonical variables are the conversion rate to sidebands  $\eta\equiv  \frac{|A_1|^2+|A_{-1}|^2}{E}$ and the relative phase $\psi\equiv \frac{\phi_1 + \phi_{-1} }{2}-\phi_{0}$, where $E\equiv |A_0|^2 + |A_1|^2+|A_{-1}|^2=V_0^2 $ is a conserved quantity of the truncated system, as well as the sideband imbalance $\chi\equiv |A_1|^2-|A_{-1}|^2$. Compared to \cite{Trillo1991c}, we consider a slightly different set of variables, more suitable to our goals. 

The Hamiltonian function is $H^{(\xi)}(\psi,\eta) \equiv \tilde \gamma E \eta(\eta-1)\cos 2 \psi + \tilde \gamma E \left(\frac{3\eta^2}{4}-\eta\right) - \beta\Omega^2 \eta$, and 
%\begin{equation}
%\begin{aligned}
% \psi' &= \pder{H^{(\xi)}}{\eta} = -\beta\Omega^2 + \tilde\gamma E \left(\frac{3\eta}{2}-1\right)	 + \tilde \gamma E (2\eta-1)\cos 2\psi\\
% \eta' &= -\pder{H^{(\xi)}}{\psi} = 2\tilde\gamma E\eta (\eta-1)\sin 2\psi,
%\end{aligned}
%\label{eq:3wave1}
%\end{equation}
\begin{equation}
\psi' = \pder{H^{(\xi)}}{\eta};\;\eta' = -\pder{H^{(\xi)}}{\psi} 
\label{eq:3wave1}
\end{equation}
where the prime denotes the derivative with respect to $\xi$. More details are found in Appendix \ref{app:3wave}. 

A final transformation to $X\equiv E\int_{0}^\xi{ \tilde \gamma (y)\ud y}$ allows us to simplify the Hamiltonian function to 
\begin{equation} 
H^{(X)}(\psi,\eta) \equiv  \eta(\eta-1)\cos 2 \psi +  \frac{3\eta^2}{4} +\alpha \eta,
\label{eq:HamiltonianX}
\end{equation} 
with $\alpha\equiv-\left[\frac{ \beta \Omega^2}{\tilde\gamma E}+1\right]=\left(\frac{\Omega}{\Omega_\mathrm {M}}\right)^2-1=-4 a_\mathrm{AB}+1$, with $a_\mathrm{AB}$ the well known parameter of the AB. Now
\begin{equation}
\dot\psi = \pder{ H^{(X)}}{\eta} ;  \;\dot\eta = -\pder{ H^{(X)}}{\psi},
\label{eq:3wave2}
\end{equation} 
where the dot denotes the derivative with respect to $X$. 

The system is modulationally unstable for $|\alpha| \le 1$, the peak gain is for $\alpha=0$, while the MI cut-off is for $\alpha=1$.

Before going on, we recall that, for constant parameters, the system \eqref{eq:3wave2} exhibits the following fixed points \cite{Trillo1991c}:
\begin{enumerate}
\item  $\tilde\psi_0=\frac{\cos^{-1} \alpha}{2},\tilde\eta_0=0$ (\textit{i.e.}, no conversion to sidebands, a center for $|\alpha|>1$, a saddle otherwise);

\item  $\tilde\psi_1=\frac{\cos^{-1} \left(-\alpha-\frac{3}{	2}\right)}{2},\tilde\eta_1 = 1$ (\textit{i.e.}, full conversion, a center for $\alpha>-\frac{1}{2}$, a saddle otherwise);

\item  $\tilde\psi_2=m\pi\; (m\in\mathbb{Z}), \tilde\eta_2=\frac{2(1-\alpha)}{7}$ (\textit{i.e.}, oscillations around finite conversion rate, which exists as a center for 
$|\alpha|\le 1$); 

\item $\tilde\psi_3 = \frac{ \pi}{2}+m\pi, \tilde\eta_3=2(1+\alpha)$ (which exists as a center for $-1<\alpha\le-\frac{1}{2}$). 
\end{enumerate}

The last case applies only to the lower half of the MI gain curve ($\Omega\le\frac{\Omega_\mathrm{C}}{2}$), where the three-wave truncation obviously breaks down and higher-order sidebands at $\pm n\Omega$, $n\in \mathbb{Z}$ are also unstable. 
 \textcolor{black}{Here, we stress that resorting to a five-wave or more truncation neither solves  this problem nor improves the description for $\frac{\Omega_\mathrm{C}}{2}\le\Omega\le \Omega_\mathrm{C}$, since the reduction to a Hamiltonian system with a small number of degrees of freedom is feasible only in the present three-wave case. Thus, it represents a mathematical complication with few practical benefits. Instead an approach based on exact NLSE solutions should be envisaged.}

\begin{figure}[hbtp]
\centering
\includegraphics[width=0.45\textwidth]{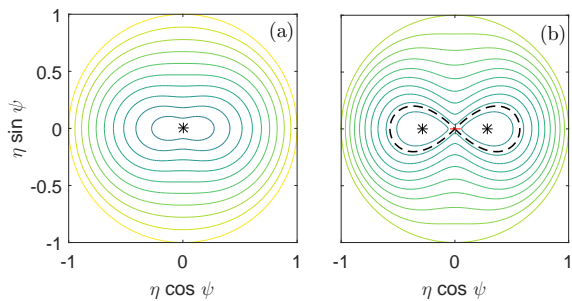}
\caption{
The two topologies of the phase-plane considered in this work. We show the level sets of the Hamiltonian $H^{(X)}$ \textcolor{black}{, Eq.~\eqref{eq:HamiltonianX},} on which a trajectory lies for constant $\alpha$. (a) $\alpha=1.25$:   the system is modulationally stable and only free rotations around $\tilde\eta_0$ (a center, marked as a black asterisk) are possible. (b) $\alpha=0$: $\tilde\eta_0$ (marked as a red plus sign) is unstable and lies on a separatrix curve, while centers $\pm\tilde\eta_2$ (marked as black asterisks) appear. In all situations, the trajectories turn anticlockwise. }
 \label{fig:phaseplanetop}
\end{figure}

The two different topologies of the phase-plane (for $\alpha\gtrless 1$) are exemplified in Fig.~\ref{fig:phaseplanetop}(a)[(b), respectively]. 
%%%%%%%%%%%%%%%%%%%%%%%%%%%%%%%
% Figure phase plane topologies
%%%%%%%%%%%%%%%%%%%%%%%%%%%%%%%%%%%%%%%%%%%%

For what follows, it is also useful to recall that, for $\alpha<1$, the trajectory emanating from $(\tilde\psi_0,\tilde\eta_0)$ is homoclinic and is referred to as a separatrix.   By direct inspection of Eq.~\eqref{eq:3wave2}, it is easy to see that trajectories always turn anticlockwise for $X>0$, irrespective of $\alpha$. This implies that the separatrix exits the origin in the second and fourth quadrants and rejoins it in the third or first, respectively. We recall also that conventionally, trajectories outside (resp.~inside) the separatrix are named period-two (resp.~one) solutions. This is apparent in Fig.~\ref{fig:phaseplanetop}(b) and corresponds to the classification of time-periodic NLSE solutions, exhibiting (or not) a phase shift \cite{Akhmediev1986}. The separatrix turns out to correspond to an AB, while the centers $(\tilde\psi_2,\tilde\eta_2)$ to the steady state dn-oidal solution \cite{Yuen1982}
It is also important  that $H^{(X)}(\tilde\psi_0,\tilde\eta_0)=0$ for all $ \alpha$. For $\alpha\ge1$, $H^{(X)}>0$ everywhere in the whole unit disk, while for $\alpha<1$, $H^{(X)}\gtrless 0 $, outside or inside the separatrix, respectively. This is obvious, by noticing that $H_ \mathrm{min}\equiv H^{(X)}(\tilde\psi_2, \tilde\eta_2)=-\frac{(1-\alpha)^2}{7}\le 0$, in its domain of existence.

In general, as the bathymetry and thus $\tilde\gamma/\beta$ vary, the change of $\alpha$ lets $H^{(X)}$ (or $H^{(\xi)}$) vary across 0, see Eq.~\eqref{eq:HamiltonianX}.
This additional degree of freedom provides the flexibility to explore the stabilization  regime we will present in the next section. 

We will refer to the results of the present section as \emph{truncated} or \emph{three-wave} model, while the numerical solutions of Eq.~\eqref{eq:DReq2} are referred to as \emph{simulations}.

\section{Stabilization over an uneven bottom}
\label{sec:adiab}

 It is well known from classical  {mechanics that} a trajectory oscillating around an elliptic fixed point keeps on following the same type of oscillatory trajectories if an internal parameter is changed adiabatically, \textit{i.e.}, the speed of variation is much smaller than the oscillation frequency  \cite{ArnoldMMCM}. For Hamiltonian system, a quantity, called the adiabatic invariant, is conserved all along the transition; this is the classical counterpart of Ehrenfest theorem in quantum mechanics. 
In order to solve our problem, we have to go beyond this result and recall the theory of autoresonance \cite{Friedland1992a,Friedland1992,Porat2013}. Two possible regimes can occur. Either the trajectory starts close to an equilibrium and a parameter is changed adiabatically, so that the adiabatic invariant is conserved; or it is forced to cross the separatrix and phase-locks in the close proximity to an equilibrium and the adiabatic invariant is not conserved. 
We explain below that the second solution is much more practical if the total transition length is constrained and for the flexibility in initial conditions.

Thus, we focus here on how to physically apply the second approach to our model. 
As we showed above in Sec.~\ref{ssec:NLMI}, our system has only elliptic fixed points for $\alpha>1$ and both unstable hyperbolic and elliptic points for $\alpha<1$. Our aim is to stabilize the trajectory around  $(\tilde\psi_2,\tilde\eta_2)$ starting from small oscillations around $\tilde\eta_0$, by varying $\alpha$.
The trajectory must thus cross the separatrix: a sign change of the Hamiltonian is associated to this transition. 

Thus, three different aspects have to be considered: (i) the initial stage where the system behaves almost linearly, (ii) the separatrix crossing stage, and (iii) small oscillations around an equilibrium adiabatically shifted towards a larger $\eta$. We describe the three successively below.

\subsection{Linear stage}
\label{ssec:linear}

We start from $\eta_0\equiv\eta(0)\ll 1$ and  $\alpha_0\equiv\alpha(0)>1$, and linearize the system of Eq.~\eqref{eq:3wave2} in order to understand its behavior when we tune the parameters to cross the bifurcation point $\alpha = 1$ from above.  By letting $R\equiv \sqrt{\eta}e^{i\psi}$, we reduce Eq.~\eqref{eq:3wave2} to
\begin{equation}
	\dot{R} = i \alpha R - i R^*,
\label{eq:linearR}
\end{equation}
which can also be obtained by linearizing the complex system reported in App.~\ref{app:3wave}, Eq.~\eqref{eq:3complexwaves}, directly. The validity of Eq.~\eqref{eq:linearR} is limited to $\eta\ll 1$; nevertheless, we can obtain some useful information about the full dynamics.  

We let $R=u+iv$ and  split Eq.~\eqref{eq:linearR} in real and imaginary part to get
\begin{equation}
\begin{aligned}
	\dot u &= -(\alpha+1) v,\\
	\dot v &= (\alpha-1) u.
\end{aligned}
\label{eq:linearRreim}
\end{equation}
If we divide these two equations term by term, we see that the solution is of the form $v^2 = C - \frac{\alpha-1}{\alpha+1}u^2$, which is either an ellipse or a hyperbola, for resp.~$\alpha\gtrless1$. For $\alpha>1$, it entails periodic oscillations, albeit, $\Lambda_\mathrm{lin}^{(0)}(\alpha)>\Lambda_\mathrm{nl}(\alpha,H^{(X)})$, defined as the periods predicted by Eqs.~\eqref{eq:linearR} and \eqref{eq:3wave2} respectively, see App.~\ref{app:Ham} and \ref{app:Hlin}.
For $\alpha<1$, Eq.~\eqref{eq:linearR} gives exponentially divergent solutions. 
For $\alpha=1$, we have a pair of straight lines $v=\pm \sqrt{C}$, \textit{i.e.}, the  horizontal semi-axis of the ellipse diverges. If $|u|\gg C$ at the same $X$, $\psi\to m\pi=\tilde\psi_2$. Thus we can define this stage as the phase-locking stage.

Finally, from Eq.~\eqref{eq:linearR}, notice that for $\alpha\gg 1$ the second term can be neglected and $R$ oscillates on a circle of radius $|R|^2\to\eta(-\infty)$; this limit gives $C = \eta(-\infty)$.

The trajectories of the full nonlinear system turn anticlockwise, so do necessarily the solutions of its linearized version [the first of Eqs.~\eqref{eq:linearRreim} clearly shows that]. In order to follow the separatrix and then cross it and approach the centers located at $ \mp\tilde\eta_2=\mp\frac{2}{7}(1-\alpha)$, $(u,v)$ are required to lie in the  second or fourth quadrant, respectively:  at $\alpha\approx 1$, we thus impose $\dot u u >0$ (or, equivalently, $u v<0$). Otherwise, the solution moves away from the elliptic fixed points and oscillates outside the separatrix. 

Let $\alpha= 1 - \Deltalpha^\mathrm{i} (X-X^*)$, with $\Deltalpha^\mathrm{i} >0$, so that at   $X=X^*>0$ we reach the MI band edge $\Omega_\mathrm{C}$.

In order to find suitable initial conditions, we resort to a local approximation in power series, shown in App.~\ref{app:linloc}. 
We conclude that, for $\alpha_0$  close to $1$ and $\psi_0\equiv \psi(0)=\pm \frac{\pi}{2}$ trajectories evolve to the correct quadrant and phase-lock to, respectively, $\pi$ or $0$, while $\psi_0= 0$ does not. 
 
A lower limit to $\Deltalpha^\mathrm i$ must be imposed.
$\alpha(X^*)=1$ gives $X^* = \frac{\alpha_0-1}{\Deltalpha^\mathrm i}$.  We require that $X^*\ll \Lambda_\mathrm{nl}/2$, \textit{i.e.}, the MI band is crossed before the system reaches the peak $\eta$. Otherwise, the trajectory would point back and could not enter the separatrix as this last appears. 
We conclude that $\Deltalpha^\mathrm i\gg \frac{2(\alpha_0-1)}{\Lambda_\mathrm{nl}}> \frac{2(\alpha_0-1)\sqrt{\alpha_0^2-1}}{\pi}$, by virtue of $\Lambda_\mathrm{nl}<\Lambda_\mathrm{lin}^{(0)}=\pi (\alpha_0^2-1)^{-\frac{1}{2}}$, as shown in App.~\ref{app:Ham} and \ref{app:Hlin}. In order to lie close but near the separatrix as it appears, we require $\eta_0\ll 1$. The Hamiltonian takes thus the value $H^{(X)}(X^*)\approx \left[1+(X^*)^2\right]v_0^2$ at the bifurcation point. 

\subsection{Intermediate regime} 
\label{ssec:interm}

Suppose that the the solution of Eq.~\eqref{eq:3wave2} behaves at $X^*$ as a trajectory close to the separatrix, Eq.~\eqref{eq:separatrix} in App.~\ref{app:Ham}, and  grows away from $\tilde\eta_0$. After an initial exponential growth, $\eta$ slows down and its growth rate starts soon decreasing. The homoclinic orbit appears at $\alpha=1$ and  expands linearly in width with decreasing $\alpha$. From Eq.~\eqref{eq:HamiltonianX}, as $\alpha \eta$ decreases, $H^{(X)}$ will change sign, thus separatrix crossing occurs. The analytic treatment to characterize the solution near this point is very involved for the system given by $H^{(X)}$ \cite{Cary1986} and does not provide hints about the dynamics of Eq.~\eqref{eq:DReq2}. 
{Nevertheless, we estimate the optimal variation of $\alpha$ and the distance at which it can be achieved by following a simpler argument, similar to what reported in Ref.~\cite{Caglioti1988}. Starting at $X^{*}$, the optimal transition is such that $H^{(X)}(X^{**})=H_\mathrm{min}$, where $X^{**}$ marks  the adiabatic stage start. In this way, the orbit reaches closely to $(\tilde\psi_2,\tilde\eta_2)$. From Eq.~\eqref{eq:HamiltonianX}, we have}
\begin{equation}
	\frac{\ud H}{\ud X} = \dot \alpha \eta \approx - \Deltalpha^\mathrm{t} \eta,
	\label{eq:hamvareq}
\end{equation}
where we assumed, as before, that $\alpha$ decreases linearly with slope $\Deltalpha^\mathrm{t}$.
{
We can thus approximately integrate Eq.~\eqref{eq:hamvareq} and write 
\begin{equation}	
H^{(X)}(X^{*}) - \frac{\Deltalpha^\mathrm{t}(X^{**}-X^{*})}{2}(\eta^*+\eta^{**})=H_\mathrm{min}. 
\label{eq:hamvar}
\end{equation}
$\eta^{*}\equiv\eta(X^*)$ is known from the linear stage above, and we take $\eta^{**}\equiv\eta(X^{**})=\tilde\eta_2$, to enforce the proximity to the center at some given distance. We thus require that \begin{equation}
\Delta\!X^\mathrm{t} \equiv X^{**}-X^{*}=\frac{\Lambda_\mathrm{nl}^{m\to1}(H^{(X)}(X^*))}{4},
\label{eq:halfperiod}
\end{equation}
\textit{i.e.}, the start and end of the intermediate stage are separated by roughly a fourth of a period of an external orbit close to the separatrix, computed at $X^*$, see App.~\ref{app:Ham}. This is justified by the fact that we have period-two solutions outside the separatrix.
By plugging these values into Eq.~\eqref{eq:hamvar}, we obtain the optimal slope for changing $\alpha$ in the intermediate stage,
\begin{equation}
 \Deltalpha^\mathrm{t} = \frac{2 H^{(X)}(X^*)}{\eta^{*} \Delta\!X^\mathrm{t}}.
 \label{eq:deltalphaI}
\end{equation}
 }
We notice that the farther we start from the separatrix, the larger the variation of $\alpha$ is required.

\subsection{Adiabatic conversion stage}
\label{ssec:adiab}
Suppose that the separatrix is crossed and, at distance $X^{**}$,  the system is close to  the center $(\tilde\eta_2,\tilde\psi_2)$ computed at the current value of $\alpha(X^{**})$. Suppose we can approximate $\alpha(X) = \alpha(X^{**}) - \Deltalpha^\mathrm{f} (X-X^{**})$. The trajectory will keep on oscillating around the  equilibrium, which in turn varies with $\alpha$, provided that an adiabaticity condition on $\Deltalpha^\mathrm f$ is satisfied. 
We estimate it by resorting to the same approach of Ref.~\cite{Porat2013}.

In App.~\ref{app:Hlin}, we discuss the general method to linearize the Hamiltonian and obtain that $(\psi,\eta)$ make small oscillations around $(\tilde\psi_2,\tilde\eta_2)$ if 
\begin{equation}
\left|\frac{2\dot\alpha}{7 \kappa_2}\right| = \left|\frac{2\Deltalpha^\mathrm{f}}{7 \kappa_2}\right| \ll 1,
\label{eq:adiabcond} 
\end{equation}
where  $\kappa_2\equiv\frac{2\sqrt{7}}{7}\sqrt{(1-\alpha)(5+2\alpha)}$ is the linearized angular frequency around the center, see App.~\ref{app:Hlin}. 

It is easy to check that $\kappa_2$ grows monotonically for $-\frac{3}{2}<\alpha<1$, thus the most stringent upper bound on $\Deltalpha^\mathrm{f}$ occurs at $X^{**}$.

\section{Numerical results}
\label{sec:results}
\subsection{Initial conditions}
\label{ssec:initialconds}
We suppose for simplicity that $\kappa$ is changed linearly all over the domain: $\alpha_0=1.56$, \textit{i.e.}, 
$\Omega=1.6\Omega_\mathrm{M}$, and  $\kappa$ varies from 2 to 5. In practice, this means a linear variation of $h$, see the magenta dashed dotted line in Fig.~\ref{fig:adiabetapsi}(a) (the scale  on the right axis). The effect on $\alpha$ is instead a faster variation in the beginning and slower after $\xi\approx200$. This is a particularly favorable situation for the locking into the elliptic fixed point, according to the previous discussion. 

The initial conditions are $\Eps(0)\equiv\Eps_0 = V_0 k(0) = 0.12$, $\eta_0=0.025$, and $\psi_0=\frac{\pi}{2}$: they are optimal according to the discussion in Sec.~\ref{ssec:linear}.

\subsection{Simulation results}
 \textcolor{black}{We solve Eq.~\eqref{eq:DReq2} by means of the adaptive 3\textsuperscript{rd}-order Runge-Kutta (RK) scheme embedding the conventional 4\textsuperscript{th}-order applied to the interaction-picture formulation \cite{Balac2013}. We use $2^{11}$ points for time-discretization, while the integration step in 
$\xi$ is adapted to keep the error below $1\times 10^{-9}$. This guarantees a short computational time (less that 30 seconds for each simulation) with a satisfactory conservation of $N$ and $P$ (deviations of less that $5\times 10^{-3}$ and $2\times 10^{-6}$, respectively).}

%%%
\begin{figure}[hbtp]
\centering
\includegraphics[width=0.42\textwidth]{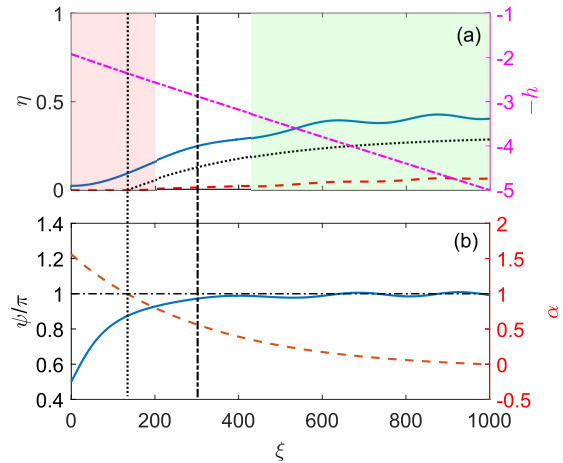}
\caption{Simulated spectral evolution over a smoothly varying depth. $\xi$ and $h$ are in units of m in the scaling discussed in the main text; the other quantities are dimensionless. (a) On the left axis, conversion efficiency $\eta$ (blue solid line), its value at the elliptic fixed point $\tilde\eta_2$ predicted by the three-wave truncation (black dotted line) computed from the local value of $\alpha$, and  the relative intensity of the second order sidebands, $\eta^{(2)}$ defined  below in the main text (red dashed line). On the right axis, we plot the bathymetry, shown as a purple dash-dotted line. (b) The evolution of the relative phase $\psi/\pi$ (blue solid lines, refers to the left axis) and the MI coefficient $\alpha$ (defined in the text, red dotted line, refers to the right axis). The crossing into the MI band is where $\tilde\eta_2$ appears; to guide the eye, panel (b) includes a thin black dashed-dotted horizontal line.  The vertical black dotted line marks the $\alpha=1$ point, while the vertical  black dash-dotted identifies the point where $H^{(\xi)}=0$ (separatrix crossing). The red (resp.~green) shaded region represents the linear (resp.~adiabatic) stage of stabilization. }
\label{fig:adiabetapsi}
\end{figure}
%%%%

In Fig.~\ref{fig:adiabetapsi}, we clearly identify the three stages described above: (i) the linear, around $\alpha=1$, where $\eta$ grows and $\psi$ approaches $\pi$ (red-shaded area); (ii) the intermediate, starting at $\xi\approx 200$, where the growth slows down, the separatrix is crossed and $\psi$ locks to $\pi$; (iii) and the adiabatic, starting at $\xi\approx 450$, where $\eta$ adiabatically follows the equilibrium up to $\eta\approx 0.5$ (green-shaded area). The residual oscillations in amplitude and phase are below 5\% and 1\% in relative terms, see blue solid lines in panels (a) and (b), respectively. The second-order sideband fraction, defined as $\eta^{(2)}\equiv\frac{|\hat V(2\Omega,\xi)|^2 + |\hat V(-2\Omega,\xi)|^2}{V_0^2}$,  represents less than 10$\%$ of the total mass $N$ [red dashed line in panel Fig.~\ref{fig:adiabetapsi}(a)]. They are generated via nonlinear processes of the sort $0 \pm \Omega \pm \Omega\to\pm2\Omega$, which are thresholdless and oscillating. They partially account for the discrepancy between the numerical solution and $\tilde\eta_2$ (black dotted line). Systematically simulations end up oscillating around a larger $\eta$ than predicted by the three-wave truncation.  \textcolor{black}{This is the main limitation of the truncated model and  mitigated in the $\Omega\to\Omega_C$ limit for small $\eta$} \cite{Trillo1991c,Armaroli2017}.

%%%%%%%%%%%%%%%%%%%%%%%%%%%%%%%%%%%%%%%%%%%%%%%%%%%
\begin{figure}[hbtp]
\centering
\includegraphics[width=0.42\textwidth]{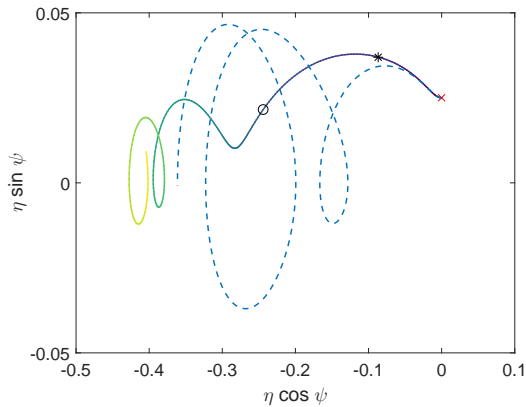}
\caption{Phase space representation of the numerical results of simulation and three-wave truncation  (solid and dashed line,	 respectively) over a smoothly varying depth. The  direction of evolution in $\xi$ is represented by the line getting a lighter hue, while the red cross denotes the initial condition and the black circle the point of separatrix-crossing where $H^{(X)}=H^{(\xi)}=0$. The asterisk corresponds to the $\alpha=1$ point.  \textcolor{black}{Notice that the horizontal and vertical axes have different scales.}}
\label{fig:adiabpp}
\end{figure}
%%%%%%%%%%%%%%%%%%%%%%%%%%%%%%%%%%%%%%%%%%%%%%%%%%%%

We compare simulations (solid line with changing hue) to the truncated model (dashed blue line) in the phase-plane, Fig.~\ref{fig:adiabpp}. In both cases $\eta$ grows, the phase is locked and the residual oscillations are very small. Notably, in the simulation the oscillations around the average are limited to less than $0.025$. 

The asterisk marks the $\alpha=1$ transition, after which the linear approximation soon breaks down.
The circle denotes instead the separatrix crossing, $H^{(\xi)}=0$. {Notice that  the trajectory turns away from the horizontal axis just after a close approach to an elliptic equilibrium [equivalent to $(\tilde\psi_2,\tilde\eta_2)$]. This occurs at $\eta=0.28$, and the phase is then locked, see Fig.~\ref{fig:adiabetapsi}(b). }

The three-wave solution (dashed line in Fig.~\ref{fig:adiabpp}) exhibits larger oscillations than the simulated ones (the horizontal and vertical scales differ much): in fact, the final value of $\kappa$ is chosen to minimize these latter.   The former meets its optimal conversion effectiveness at $\kappa(\xi =1000)\approx 5.5$, which combines the fast locking condition with the adiabatic following of the center: $\Deltalpha^\mathrm{i}\gg 0.5$ at $\xi=0$, while $\Deltalpha^\mathrm{f}\ll 5$ at $\xi=450$. For such a $\kappa$, the simulation turns out to oscillate more, which we explain by the faster displacement of the elliptic fixed point of the NLSE compared to $\tilde\eta_2$, \textit{i.e.}, {the conditions  \eqref{eq:deltalphaI} and \eqref{eq:adiabcond} are stricter for the NLSE than for the truncated model}. This is again inherent to the three-wave approximation.

In principle there is no limit on how large the fraction of $N$ can be funneled into $\eta$. 
The  physical range of $\tilde\gamma/\beta$ is nevertheless limited, see Fig.~\ref{fig:DRparams}(a). 
Finally, once the total length of the system is constrained, $\Deltalpha^\mathrm f$ is bounded from below. The condition of Eq.~\eqref{eq:deltalphaI} looks quite more stringent, but we verified numerically that, provided the separatrix is crossed, the behavior is very similar to the optimal one: the blue dashed line in Fig.~\ref{fig:adiabpp}, pertaining to the three-wave model, shows indeed the typical solution.

%%%%%%%
% Potential well
\begin{figure}[hbtp]
\centering
\includegraphics[width=0.42\textwidth]{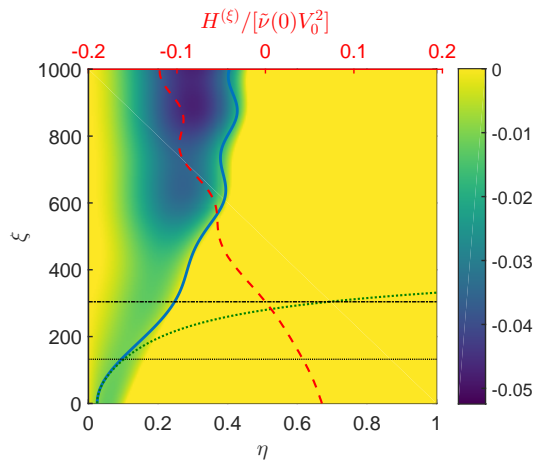}
\caption{ \textcolor{black}{Simulated} evolution of  $\eta$ in the $\xi$-dependent potential well \textcolor{black}{, explicitly derived for the three-wave model in App.~\ref{app:Ham}}. The color-map represents the  values of the function $-W(\eta)$ (the darker the deeper negative, yellow regions are classically inaccessible). We let $H^{(\xi)}$ vary in $\xi$ and the potential well is recomputed accordingly for each point in the evolution \textcolor{black}{, by replacing values extracted from simulations}. The blue solid line corresponds to the numerical solution of Eq.~\eqref{eq:DReq2}, while the dark green dotted curve represents the solution of Eq.~\eqref{eq:linearR}, $|R|^2$. The red dashed line shows the values of the three-wave Hamiltonian  \textcolor{black}{ 	calculated from the solution of Eq.~\eqref{eq:DReq2},} mapped on the top axis. Like in Fig.~\ref{fig:adiabetapsi}, the black dotted and dash-dotted lines represent the distances at which $\alpha=1$ and $H^{(\xi)}=0$. }
\label{fig:adiabpotential}
\end{figure}
%%%%%%%%%%%%%%%%%%%%%%%%%%%%%%%%%%%%%%%

A third alternative representation is available. Recall that Eqs.~\eqref{eq:3wave1}-\eqref{eq:3wave2} are equivalent to a particle moving in a potential well, as explained in App.~\ref{app:Ham}. Notice that the potential well $W(\eta)$ depends on the initial value of $H^{(X)}$, thus its minima do not correspond to equilibria, in general. We let $H^{(X)}$ vary and  update it at each integration step  \textcolor{black}{by replacing values of $\eta$ and $\psi$ extracted from simulations}, according to the $X$-dependent expression Eq.~\eqref{eq:HamiltonianX}, see App.~\ref{app:Ham} for more details. In Fig.~\ref{fig:adiabpotential} we show  the map of the accessible values of the potential well $-W(\eta)\le 0$:  this is very shallow and narrow at the beginning (where $\alpha>1$), then it becomes broader and deeper. Again, we see that the linear approximation, dark green dotted line, diverges at $\xi\approx200$. After this linear stage, the well smoothly widens and deepens; $H^{(\xi)}$ changes sign at $\xi\approx 300$.
In the last stage, from $\xi\approx 450$, the potential well gets deeper and deeper and the results of the simulation (blue solid line) is clearly trapped into it, as expected by the adiabatic following of the elliptic fixed point, proven above, and in spite of the systematic difference with the three-wave results.

%%%%%%%%%%%%%%%%%%%%%%%%%%%%%%%%%%%%%%%%%%%%%%%%%
\begin{figure}[hbtp]
\centering
\includegraphics[width=0.42\textwidth]{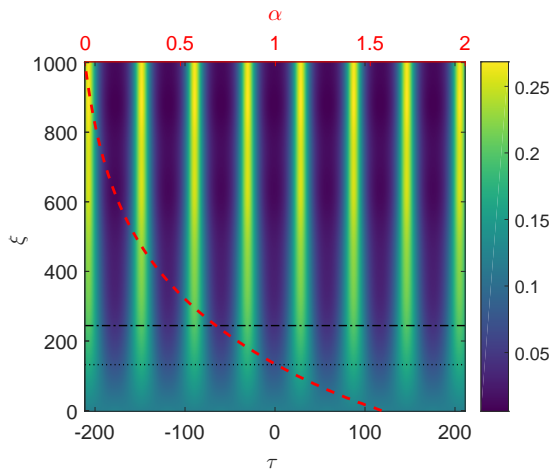}
\caption{Color-map of the space-time evolution of steepness $kU$, with $U$ the solution of Eq.~\eqref{eq:DReq1}, over an uneven bottom with linear slope. $\xi$ and $\tau$ are in units of $\mathrm{m}$ and $\mathrm{m}^\frac{1}{2}$, respectively, as discussed in the text. The red dashed line shows the variation of $\alpha$ as a function of $\xi$ and is mapped on the additional abscissa on the top. The two horizontal black lines (dotted and dashed-dotted) correspond to their counterpart in Fig.~\ref{fig:adiabpotential}.}
\label{fig:adiabspacetime}
\end{figure}
%%%%%%%%%%%%%%%%%%%%%%%%%%%%%%%%%%%%%%%

Finally, a further limitation inherent to  water waves is that nonlinear effects cannot be increased arbitrarily,  {because they scale as the generalized Ursell number  \cite{Osborne1994}, which is proportional to $\Eps$ in the deep water limit}. An AB envelope peaks at roughtly two-to-three times the background amplitude, and  wavebreaking occurs if $\varepsilon\gtrsim 0.4$ \cite{babanin2011breaking}. The physical soundness of our approach is confirmed by representing the evolution of $\varepsilon$ attained by $U$---the solution of Eq.~\eqref{eq:DReq1}, which represents the envelope of physical surface elevation, see Fig.~\ref{fig:adiabspacetime}. We notice that the proposed stabilization technique almost completely suppresses oscillations of $U$; this reflects in negligible $\Eps$ overshoots, never larger than 0.3, which guarantees that the train of pulses will not break. 

\subsection{A glimpse into a physical realization in hydrodynamics}
\label{ssec:numbers}

In the previous section we use a quite conservative set of parameters, in order to assure the validity of the NLSE and the non-breaking of the wavetrain. 
The question arises if the stabilization can be  achieved in a laboratory setting.

We consider a 100 m long wave tank \textcolor{black}{, which is feasible in state of the art hydrodynamic facilities. As our $\xi=1000$-long domain reduces to this length, all the other quantities presented in Sec.~\ref{ssec:initialconds} are automatically rescaled. We obtain a carrier frequency $f=1.58$ Hz, the corresponding depth values are then $h(x=0) = 20$ cm and $h(x=100\, \mathrm{m})=50 $ cm, the local wavelength varies from $\lambda(x=0) = 63$ cm to $\lambda(x=100 \,\mathrm{m})=157 $ cm, and the sideband detuning converts to $\Delta\! f = 0.17$ Hz. Finally, the maximum wave amplitude is estimated to vary from the initial 1.2 cm to 6.5 cm at the end of the wave tank. }

This is an idealization, because damping occurs. From Eq.~\eqref{eq:effectiveNL}, we notice that the shoaling partially compensates dissipation. Nevertheless, shoaling becomes negligible for larger $\kappa$, while the wave field keeps on damping exponentially: mathematically it is impossible to have exact compensation because $c_\mathrm{g}$ decreases with $\kappa$. Moreover, we showed above that phase-locking is kept only if $\alpha$ is changed slowly. Thus, it is not possible to simply choose an arbitrarily large $\kappa$ so that $\tilde\gamma$ reaches the same values of the undamped case; in fact, this would lie outside of the accessible parametric range because, for $\kappa>5$, $\gamma$ is almost constant, see Fig.~\ref{fig:DRparams}.  

The pulse train is thus meta-stable: for large enough damping, the separatrix will be crossed again, a period-two solution will be observed, and eventually the wave will vanish completely \cite{Kimmoun2016,Soto-Crespo2017}. 

The analytic treatment is as involved as the one required to describe the second stage of the stabilization. 

We found from \textit{simulations} that keeping every parameter as before a total loss of 20\% can be tolerated. For the wave tank length specified above, this corresponds to $\nu \approx 2 \times 10^{-3} \,\mathrm{m}^{-1}$, which is reasonable when the effect of sidewall dissipation is taken into account \cite{HUNT1952a}.

\section{Conclusions}
\label{sec:concl}
In this work we studies the nonlinear stage of evolution of modulational instability in surface water waves over a water body of gradually increasing depth. We showed that this stage can be stabilized and results in a uniform train of pulses on a background. The initial condition does not need to be restricted to an exact NLSE solution (e.g., an Akhmediev breather), but just a harmonic perturbation with a given small amplitude. 

Based on a three-wave truncation, we studied how a linear depth change naturally leads to a virtually frozen state (which can be considered close to dn-oidal solution of the NLSE), provided that suitable initial conditions, i.e., frequency lying just outside the instability margin and with a relative phase facilitating separatrix crossing, are chosen. 

Within these restrictions, still a wide range of carrier frequencies and depth variation is compatible with stabilization, even in spite of the unavoidable viscous damping.

Although the flexibility available to vary parameters in the hydrodynamics of surface water waves is much less than in other physical systems, such as optical fibers, our results will help clarify the possibility to dynamically control the breathing evolution of water wave-packets and to understand the impact of bathymetry on the persistence (or lifetime) of rogue waves. 

 {Finally, we emphasize that the dn-oidal solution exhibits a specific stationary spectral profile, where sidebands are all phase-locked and where sidebands are in a given ratio among one another \cite{Magnani2019}.
 While we find our approach more general and physically transparent, the transformation between general solutions of the NLSE will be the subject of future studies.}

\begin{acknowledgements}
	We acknowledge the Swiss national Science foundation (SNF grant 200020 175697) and the University of Sydney--University of Geneva Partnership collaboration award.

\end{acknowledgements}

\appendix

\section{Coefficients of Eq.~\eqref{eq:DReq1}}
\label{app:VDcoeffs}

We recall here the coefficients derived in \cite{Djordjevic1978}. Some misprints were present in the original manuscript. The correct version is  found also in \cite{Chiang2005}. 

The dispersion coefficient reads as
\begin{equation}
\beta\equiv-\frac{1}{2\omega c_\mathrm{g}}\left[1-\frac{gh}{c^2_\mathrm{g}}(1-\kappa\sigma)(1-\sigma^2)\right].
\end{equation}
with fixed $\omega=\sqrt{g k \sigma}$, $\sigma\equiv\tanh kh$. Recall the expression of group velocity $c_\mathrm{g} \equiv \pder{\omega}{k}= \frac{g}{2\omega}\left[\sigma + \kappa(1-\sigma^2)\right]$, while the phase velocity $c_\mathrm{p}=\frac{\omega}{k}=\sqrt{\frac{g\sigma}{k}}$.

The nonlinear coefficient reads as 
\begin{equation}
\begin{aligned}
\gamma\equiv&\frac{\omega k^2}{16\sigma^4 c_g}\left\{9-10\sigma^2+9\sigma^4-\right.
\\
&\left.\frac{2\sigma^2c^2_g}{gh-c_g^2}\left[4\frac{c^2_p}{c^2_g}+4\frac{c_p}{c_g}(1-\sigma^2)+\frac{gh}{c^2_g}(1-\sigma^2)^2\right]       \right\}.
\end{aligned}
\end{equation}

These expressions  are simplified by using a more natural system of units where we let $g=1$ (without dimension). Time and speeds are in units of $[m^{\frac{1}{2}}]$, frequencies in unit of  $[m^{-\frac{1}{2}}]$. Given the simple scaling of coefficients, we can assume $\omega=1$ throughout the paper, without loss of generality.
 
Finally, the shoaling coefficient reads as
 \begin{equation}
 \mu_0\equiv\frac{1}{2 \omega c_g}\frac{\ud\left[ \omega c_g\right]}{\ud \kappa}=\frac{(1-\sigma^2)(1-kh \sigma)}{\sigma + kh\left(1- \sigma^2\right)} 
 \end{equation}

\section{Three-wave truncation: from complex to real variables}
\label{app:3wave}

Let us substitute the Ansatz $V(\xi,\tau)= A_0(\xi) + A_1(\xi) e^{i \Omega \tau} + A_{-1}(\xi)e^{-i\Omega \tau}$ in Eq.~\eqref{eq:DReq2}. By retaining only the terms oscillating at the frequencies $0$ and $\pm\Omega$, we obtain
\begin{equation}
\begin{aligned}
	iA_0' =&\tilde\gamma  (|A_0|^2 + 2 |A_1|^2 + 2 |A_{-1}|^2 )A_0 \\&+  2\tilde\gamma   A_{1}A_{-1}A_0^*\\
	iA_1' =& \beta\Omega^2 A_1 +\tilde\gamma  (|A_1|^2 + 2 |A_0|^2 + 2 |A_{-1}|^2 )A_1 \\ &+ \tilde\gamma   A_{-1}^*A_0^2\\
	iA_{-1}' =& \beta\Omega^2 A_{-1} +\tilde\gamma  (|A_{-1}|^2 + 2 |A_0|^2 + 2 |A_{1}|^2 )A_{-1} \\&+  \tilde\gamma   A_{1}^*A_0^2
\end{aligned}
\label{eq:3complexwaves}
\end{equation}

Then, let $A_n = \sqrt{\zeta_n}\exp i \phi_n$, with $\zeta_n$ and $\phi_n$ real functions. 
By replacing these variables in Eq.~\eqref{eq:3complexwaves}, we notice that $\phi_n$ appear only in the relative phase $\psi$ defined in the main text. Moreover, it is easy to observe that the total intensity $E\equiv \zeta_0 + \zeta_{1} + \zeta_{-1}$ as well  as the sideband imbalance $\chi\equiv\zeta_1-\zeta_{-1}$ are conserved. It is thus practical to define $\eta$ as in the main text, so that $\eta\in[0,1]$.

\section{Hamiltonian formalism and some analytical results}
\label{app:Ham}

Some trivial algebra allows one to rewrite Eq.~\eqref{eq:3complexwaves} as
\begin{equation}
\begin{aligned}
	\psi' &=  -\beta \Omega^2 + \tilde\gamma E \left[\frac{3\eta}{2}-1\right]			+\tilde\gamma E S\cos 2\psi \left[1+\frac{\eta(\eta-1)}{S^2}\right]
	\\
	\eta' &= 	2\tilde\gamma E S(\eta-1)\sin2\psi,
\end{aligned}
\label{eq:3wavesetapsi}
\end{equation}
with $S = \left[(\eta-\tilde\chi)(\eta+\tilde\chi)\right]^{\frac{1}{2}}$ and $\tilde\chi=\frac{\chi}{E}$.
Eq.~\eqref{eq:3wavesetapsi}  is integrable. Further, by transforming to the variable $X$ (defined in the main text), the system \eqref{eq:3wavesetapsi} can be cast as a one d.o.f. integrable system, with Hamiltonian function 
\begin{equation}
H^{(X)}(\psi,\eta)= S(\eta-1) \cos2\psi + \alpha \eta + \frac{3}{4}\eta^2
\label{eq:hamiltonianfull}
\end{equation}

A simple transformation allows us to derive a separable equation of the form 
\begin{equation}
	\dot\eta^2 = W(\eta),
	\label{eq:resolvent}
\end{equation}
with 
\begin{equation}
\begin{aligned}
	W(\eta;H,\tilde\chi) &= 4\left[\frac{7}{16}\eta^4 - \left(2 +\frac{3}{2}\alpha\right)\eta^3 \right. \\	
	&\left.+\left(1-\alpha^2-\tilde\chi^2 +\frac{3H}{2}\right)\eta^2 \right. \\
	&\left.+ 2\left(\alpha H+\tilde\chi^2\right) \eta -H^2 - \tilde\chi^2\right],
\end{aligned}
\label{eq:3wavepotential1}
\end{equation}
where $H = H^{(X)}(X=0)$ is the value of the Hamiltonian determined by the initial conditions of the problem.  Eq.~\eqref{eq:resolvent} is in the conventional $\frac{\dot\eta^2}{2} = E-V(\eta)$ form, which allows one to solve any one d.o.f. mechanical system, $W(\eta)$ plays the role of the potential well $V(\eta)$ conventionally used for textbook Hamiltonian systems. 

In the main text, we discuss only the case of $\tilde\chi=0$. 
The potential in this case reads as
\begin{equation}
\begin{aligned}
	W(\eta;H,\tilde\chi=0) &= 4\left[\frac{7}{16}\eta^4 - \left(2 +\frac{3}{2}\alpha\right)\eta^3 \right. \\	
	&\left.+\left(1-\alpha^2 +\frac{3H}{2}\right)\eta^2 + 2\alpha H \eta -H^2\right],
\end{aligned}
\label{eq:3wavepotential2}
\end{equation}
The zeros of a quartic potential can be calculated analytically and determine the dynamics of the system. 

We can solve Eq.~\eqref{eq:resolvent} in terms of Jacobi elliptic functions, see \cite{Cappellini1991} for the detailed method. Among its solutions, the separatrix---the homoclinic orbit connecting the origin to itself, which exists for $-\frac{1}{2}<\alpha<1$ and on which $H=0$---can be written in terms of elementary functions and is useful to our goals. It reads
\begin{equation}
\eta(X) = \frac{2(1-\alpha^2) }{(2+\frac{3}{2}\alpha)+(\frac{3}{2}+{2}\alpha) \cosh \left[2\sqrt{1-\alpha^2}(X -X_0) \right]}.	
\label{eq:separatrix}
\end{equation}
This means that at $X_0$ it has a peak $\eta_\mathrm{S}\equiv\frac{4(1-\alpha)}{7}=2\tilde\eta_2$. 

The period of oscillations can also be  computed as
\begin{equation}
 \Lambda_\mathrm{nl} = 2 \int_{\eta_-}^{\eta_+}{\frac{\ud \zeta}{\sqrt{W(\zeta)}} },
 \label{eq:nlperiod}
\end{equation}
where $\eta_{\pm}$ are two classical turning points, namely $W(\eta_{\pm})=0$.

For the present study, in order to compare to the results of App.~\ref{app:Hlin}, we just mention that for $H>0$, $W(\eta)$ has zeros $\{a,c\}=2(1+\alpha\pm\sqrt{(1+\alpha)^2- H})$, $\{b,d\}=\frac{2}{7}(1-\alpha\pm\sqrt{(1-\alpha)^2 + 7 H})$, with $a>b>c>d$. 
In Eq.~\eqref{eq:nlperiod} we use $\eta_-=c$ and $\eta_+=b$, to obtain $\Lambda_\mathrm{nl} = \frac{4}{\sqrt{7}}p K(m)$, where $K(m)$ is the complete elliptic integral of the first kind of parameter $m = \frac{(b-c)
(a-d)}{(a-c)(b-d)}$, $p=2\left[(a-c)(b-d)\right]^{-\frac{1}{2}}$. 
If $\alpha>1$, we have a period-one solution around $(\tilde\psi_0,\tilde\eta_0)$ and we approximate $\Lambda_\mathrm{nl}\approx\frac{\pi }{\sqrt{\alpha ^2-1}}-\frac{3 \left(\pi  \left(2 \alpha ^2+4 \alpha +1\right)\right) H}{4 \left(\alpha ^2-1\right)^{5/2}}<\Lambda_\mathrm{lin}^{(0)}=\frac{\pi }{\sqrt{\alpha ^2-1}}$, derived in App.~\ref{app:Hlin}. Thus, for $\alpha>1$ the nonlinear period is always less than the linearized approximation around $(\tilde\psi_0,\tilde\eta_0)$.
For $\alpha<1$, we have period-two solutions, instead. For $m\to 1$ we find the period close to the separatrix:  it diverges logarithmically as
$\Lambda_\mathrm{nl}^{m\to1} = \frac{2}{\sqrt{7}}p \log \frac{16}{1-m}$.

For $\alpha<1$, we consider period-one oscillations inside the separatrix, around $(\tilde\psi_2,\tilde\eta_2)$. Now, $H<0$, we redefine the roots of $W(\eta)$ as  $\{a,d\}=2(1+\alpha\pm\sqrt{(1+\alpha)^2- H})$, $\{b,c\}=\frac{2}{7}(1-\alpha\pm\sqrt{(1-\alpha)^2 + 7 H})$, with $a>b>c>d$. Integration of Eq.~\eqref{eq:nlperiod} from $\eta_-=c$ and $\eta_+=b$ gives the same expression as above, \textit{mutatis mutandis}. Again, at first order this coincides with the period $\Lambda_\mathrm{lin}^{(2)}$ derived in App.~\ref{app:Hlin}.

\section{Linearized orbits around centers}
\label{app:Hlin}

In the main text we recalled that for $\alpha>1$, $\eta=\tilde\eta_0 = 0$ is a center, and for $\alpha<1$ we have a pair of centers on the real axis $\pm\tilde\eta_2=\frac{2(1-\alpha)}{7}$. If a trajectory starts close to one of them, it continues oscillating. This oscillations can be characterized by linearizing the $X$-dependent Hamiltonian 
\begin{equation}
H^{(X)}(\psi,\eta,X) \equiv  \eta(\eta-1)\cos 2 \psi +  \frac{3\eta^2}{4} +\alpha(X) \eta.
\label{eq:XDHamilltonian1}
\end{equation}

Define $\delta\eta \equiv \eta-\tilde\eta_\mathrm{C}$ and $\delta_\psi\equiv \psi-\tilde\psi_\mathrm{C}$; subscripts C are used to denote a generic center. The evolution is derived from the linearized Hamiltonian 
\begin{equation}
\begin{gathered}
	\bar{H}(\delta\!\psi,\delta\!\eta,X) =  \left.H^{(X)}\right\rvert_{\tilde\eta_2} + \frac{1}{2} \left. \pdern{H^{(X)}}{\psi}{2}\right\rvert_{\tilde\eta_\mathrm{C}} \delta\!\psi^2
	\\+  \left.\frac{\partial^2 H}{\partial \psi\partial \eta}\right\rvert_{\tilde\eta_\mathrm{C}} \delta\!\psi\delta\!\eta
		+\frac{1}{2}  \left.\pdern{H^{(X)}}{\eta}{2}\right\rvert_{\tilde\eta_\mathrm{C}} \delta\!\eta^2 + \mathcal{O}(\eta^3,\psi^3)
\end{gathered}
\label{eq:XDHamiltonianLin}
\end{equation}

By assuming that the fluctuations evolve much faster than the equilibrium does, the equations of motion can be written as
\begin{equation}
\dot {\delta\!\eta} = -\pder{\bar H}{\delta\!\psi} - \frac{\ud \tilde\eta_\mathrm{C}}{\ud X},\;  \dot {\delta\!\psi} = \pder{\bar H}{\delta\!\eta} - \frac{\ud \tilde\psi_\mathrm{C}}{\ud X}.
\label{eq:adiabaticevol}
\end{equation}
 
First consider $(\tilde\psi_0,\tilde\eta_0)$, for $\alpha>1$. We have, obviously, $\frac{\ud \tilde\eta_\mathrm{0}}{\ud X}= \frac{\ud \tilde\psi_\mathrm{0}}{\ud X}=0$, 
$\left.\pdern{H}{\psi}{2}\right\rvert_{\tilde\eta_0} = 0$, $\left.\frac{\partial^2 H}{\partial \psi\partial \eta}\right\rvert_{\tilde\eta_\mathrm{0}}=2\sqrt{1-\alpha^2}$, and $ \left.\pdern{H}{\eta}{2}\right\rvert_{\tilde\eta_0} = \frac{3}{2} + 2\alpha$.

From Eq.~\eqref{eq:adiabaticevol}, we obtain the linear oscillator equation $\ddot{\delta\!\psi}+\kappa_0^2\psi = 0$, with $\kappa_0 \equiv 2\sqrt{\alpha^2-1}$, from which we see that the period of spatial oscillations is approximately 
$\Lambda_\mathrm{lin}^{(0)}=\pi \left[\alpha^2-1\right]^{-\frac{1}{2}}$.

Consider then $(\tilde\psi_2,\tilde\eta_2)$. Now, $\frac{\ud \tilde\eta_\mathrm{2}}{\ud X}= -\frac{2}{7}\dot{\alpha}$, $\frac{\ud \tilde\psi_\mathrm{2}}{\ud X}=0$, 
 $ \left.\pdern{H}{\psi}{2}\right\rvert_{\tilde\eta_2} = -4 \tilde\eta_2(\tilde\eta_2-1)= \frac{8}{49}(1-\alpha)(5+2\alpha)$,  $\left.\frac{\partial^2 H}{\partial \psi\partial \eta}\right\rvert_{\tilde\eta_\mathrm{2}}=0$, and $\left.\pdern{H}{\eta}{2}\right\rvert_{\tilde\eta_2} = \frac{7}{2}$;

Eq.~\eqref{eq:adiabaticevol} leads to the forced harmonic oscillator 
\begin{equation}
\ddot{\delta\!\psi} + \kappa_2^2 \delta\!\psi = \dot\alpha,
\label{eq:forcedoscillator}	
\end{equation}
 where $\kappa_2\equiv\frac{2\sqrt{7}}{7}\sqrt{(1-\alpha)(5+2\alpha)}$. The spatial period is thus $\Lambda_\mathrm{lin}^{(2)}=\sqrt{7} \pi \left[(1-\alpha)(5+2\alpha)\right]^{-\frac{1}{2}}$.

Following \cite{Porat2013}, we solve Eq.~\eqref{eq:forcedoscillator} for $\delta\!\psi(0)=\delta\!\eta(0)=$ and assuming constant $\dot\alpha$ to obtain $\delta\!\psi = \frac{2\dot\alpha}{\kappa_2^2}\sin^2 \frac{\kappa_2 X}{2}$ and $\delta\!\eta = \frac{2\dot\alpha}{7\kappa_2}\sin\kappa_2 X$. This allows us to impose the adiabatic condition on $\dot\alpha$, see Eq.~\eqref{eq:adiabcond} in the main text.

\section{Local solution of Eq.~\eqref{eq:linearRreim}}
\label{app:linloc}

We consider the behavior of Eq.~\eqref{eq:linearRreim} around $\alpha=1$. At this point the evolution of $v$ has an essential singularity; nevertheless, we numerically find that the solution is regular and this fact is the key to phase-locking.

Let $\alpha= 1 - \Deltalpha^\mathrm i (X-X^*)$, with  $\Deltalpha^\mathrm i >0$. At distance   $0<X=X^*\ll\Lambda_\mathrm{lin}^{(0)}$ we reach the MI band edge $\Omega_\mathrm{C}$.

We look for a solution of Eq.~\eqref{eq:linearRreim} of the kind $v(X)=\sum_{n=0}^{\infty}a_n (X-X^*)^n$ and $u(X)=\sum_0^{\infty}b_n (X-X^*)^n$, and obtain, by trivial algebra,
 \begin{equation}
 \begin{aligned}
 	v(X) &= a_0-\frac{\Deltalpha^\mathrm i b_0}{2}(X-X^*)^2 + \frac{2\Deltalpha^\mathrm i a_0}{3}(X-X^*)^3+\ldots \\
 	u(X) &= b_0-{2 a_0}(X-X^*) +\frac{\Deltalpha^\mathrm i a_0}{2}(X-X^*)^2 \\
 	& +\frac{\Deltalpha^\mathrm i b_0}{3}(X-X^*)^3+\ldots, 
 \end{aligned}
 \label{eq:linearRseries}
 \end{equation}
 with $a_0$ and $b_0$  arbitrary constants. 
From these expressions, it is easy to verify that the phase-locking conditions stated in the main text---$uv<0$ for $X\gtrsim X^*$---are equivalent to $a_0b_0<0$. Indeed, 
$v(0)=a_0$ is a positive minimum (negative maximum), for $b_0\lessgtr 0$. The other extremum of $v$  if for $X-X^*=\frac{b_0}{2a_0}<0$. As far as $u$ is concerned, it has a single maximum (minimum) at $X-X^*=\frac{2}{\Deltalpha^\mathrm i}>0$.

Now, we can find the best initial conditions for achieving phase-locking.
If $u(0)=0$ and $v(0)=v_0$, we obtain $a_0\approx v_0$ and $b_0 \approx- v_0 X^*$: $a_0b_0<0$. If, instead,  $u(0)=u_0$ and $v(0)=0$, we obtain $a_0\approx \Deltalpha^\mathrm i u_0 (X^*)^2$ and $b_0 \approx u_0 $: $a_0b_0>0$ and the conditions for crossing the separatrix are violated.  The approximation signs are valid if $\Deltalpha^\mathrm i (X^*)^n\ll 1$, for $n\geq 1$. Finally, notice that  $\tan\psi(X^*)=\frac{a_0}{b_0}$, thus $\tan\psi\to 0$, \textit{i.e.} phase-locked trajectories, only for the former condition.

%%%%
% Figure: solutions of R for fixed alpha
\begin{figure}[hbtp]
\centering
\includegraphics[width=0.4\textwidth]{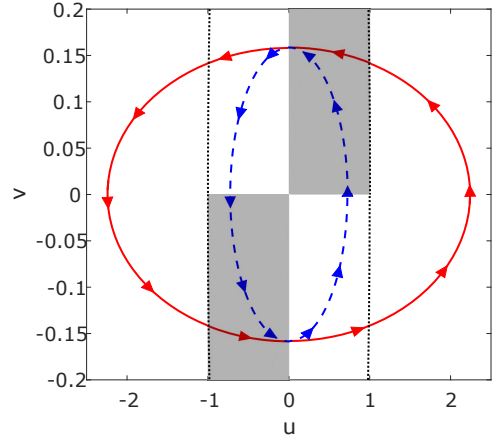}
\caption{Solutions of Eq.~\eqref{eq:linearR} for fixed $\alpha$. The blue dashed line is for $\alpha = 1.1$, the red solid line for $\alpha=0.01$. The dotted black line represents the unit circle (the axes have different scales). Only the unshaded regions inside the unit circle correspond to trajectories leading to phase-locking.}
\label{fig:Rellipses}
\end{figure}
%%%%%

We graphically illustrate these results in Fig.~\ref{fig:Rellipses}. We show two different trajectories $v^2 +  \frac{\alpha-1}{\alpha+1}u^2 = C$, with $C = \eta_0 = 0.025$ and $\alpha\in\{1.1,1.01\}$. The orbits continuously move from one ellipse to another of bigger horizontal semi-axis. In order for the initial conditions to permit phase-locking, we require that they cross into the second or fourth quadrants before $\alpha=1$. This intuitively justifies also the lower bound on $\Deltalpha^\mathrm{i}$ discussed in the main text.

An alternative local solution is to consider a second order equation for $u$, which reads
\begin{equation}
\ddot u + \frac{\dot \alpha}{\alpha+1}\dot u + (\alpha^2-1)u=0,
\label{eq:linearRe2}
\end{equation}
and gives the same solutions of Eq.~\eqref{eq:linearRseries}.

%Alternatively, Eq.~\eqref{eq:linearRe2} can be used to estimate the asymptotic trend of $R$. We observe that for constant $\alpha$, $u\sim e^{(1-\alpha^2)^\frac{1}{2}X}$, thus we can solve for $u(X) = w(X)e^{(1-\alpha^2)X}$, to obtain 
\end{document}